\documentclass[pra,aps,twocolumn,superscriptaddress,showpacs]{revtex4}

\usepackage{graphicx}

\begin{document}

\title{Entanglement between an auto-ionization system and a neighbor atom}

\author{Anton\'{i}n Luk\v{s}}
\affiliation{RCPTM, Joint Laboratory of Optics of Palack\'{y}
University and Institute of Physics of Academy of Science of the
Czech Republic, Faculty of Science, Palack\'{y} University, 17.
listopadu 12, 77146 Olomouc, Czech Republic.}
\author{Jan Pe\v{r}ina~Jr.} \affiliation{Institute of Physics of
Academy of Sciences of the Czech Republic, Joint Laboratory of
Optics of Palack\'{y} University and Institute of Physics of
Academy of Science of the Czech Republic, 17.listopadu 12, 772 07
Olomouc, Czech Republic.} \email{perinaj@prfnw.upol.cz}
\author{Wieslaw Leo\'nski}
\affiliation{Quantum Optics and Engineering Division, Institute of
Physics, University of Zielona G\'ora, Prof.~Z.~Szafrana 4a,
65-516 Zielona G\'ora, Poland.}
\author{Vlasta Pe\v{r}inov\'{a}}
\affiliation{RCPTM, Joint Laboratory of Optics of Palack\'{y}
University and Institute of Physics of Academy of Science of the
Czech Republic, Faculty of Science, Palack\'{y} University, 17.
listopadu 12, 77146 Olomouc, Czech Republic.}

\begin{abstract}
Entanglement between two electrons belonging to an auto-ionization
system and a neighbor two-level atom produced by the dipole-dipole
interaction is studied. The entanglement is quantified using the
quadratic negativity of a bipartite system including the continuum
of states. Suitable conditions for the generation of highly
entangled states of two electrons are revealed. Internal structure
of the entanglement is elucidated using the spectral density of
quadratic negativity.
\end{abstract}




\maketitle

\section{Introduction}

Ionization is a process in which an electron is transferred from
its bound discrete state into a continuum of free states, e.g., by
interacting with an optical field. In a stationary optical field,
an electron at an atom gradually leaves its bound state and moves
into an ionized free state \cite{Matulewski2003}. This process is
irreversible. It can be utilized for the generation of entangled
electron states that are stable in time. The time-dependent
entanglement among bound electrons can easily be generated in
reversible interactions (Coulomb interaction, dipole-dipole
interaction). The irreversible ionization can subsequently
'freeze' it and provide this way the stability in time. We
demonstrate this approach on the simplest model of two atoms, one
of which allows the electron ionization.

It is well known that the process of ionization is strongly
influenced by the presence of additional discrete excited states
(auto-ionization levels). They considerably modify the long-time
photoelectron ionization spectra (for an extended list of
references, see, e.g.
\cite{Agarwal1984,Leonski1987,Leonski1988,Leonski1988a}). There
even might occur Fano zeros
\cite{Fano1961,Rzazewski1981,Lambropoulos1981} in the spectra of
isolated auto-ionization systems due to the mutual interference of
different ionization paths. The interaction of an auto-ionization
system with neighbor atoms leads to the presence of dynamical
zeros
\cite{Luks2010,PerinaJr2011,PerinaJr2011a,PerinaJr2011b,PerinaJr2011c}
that occur periodically in time. Ionization spectra contain useful
information about bound states of an atom and that is why they
have widely been studied experimentally \cite{Journel1993}.
Auto-ionization systems have also been found useful as media
exhibiting electromagnetically-induced transparency and slowing
down the propagating light \cite{Raczynski2006}. The ionization
process is also sensitive to quantum properties of the optical
field \cite{Leonski1993}.

Here, we consider two atoms in a stationary optical field that
moves electrons from their ground states into excited or ionized
states. Electrons in their excited states mutually interact by the
dipole-dipole interaction \cite{Silinsh1994}. This creates quantum
correlations (entanglement) between two electrons. Whereas one
electron remains in a bound state, the second one is allowed to be
ionized. We pay attention both to the temporal entanglement
formation \cite{Bouwmeester2000,Nielsen2000} and its long-time
limit. The quadratic negativity of a bipartite system generalized
to the continuum of states is used to quantify the entanglement.
We show that highly entangled states can be reached in a wide area
of parameters characterizing the system of two atoms.

The paper is organized as follows. A semiclassical model of the
system under consideration is described in Sec.~II together with
its the most general solution. The formula for negativity as a
measure of entanglement in a bipartite system with the continuum
of states is derived in Sec.~III and compared with quantum
discord. The spectral density of quadratic negativity is
introduced in Sec.~IV. The dynamics of entanglement as well as its
long-time limit are discussed in Sec.~V. The spectral entanglement
is analyzed in Sec.~VI. Conclusions are drawn in Sec.~VII.
Appendix A is devoted to an alternative derivation of the formula
for negativity.

\section{Semiclassical model of optical excitation of
an auto-ionization atom interacting with a neighbor atom}

\begin{figure}  
 \includegraphics[scale=0.7]{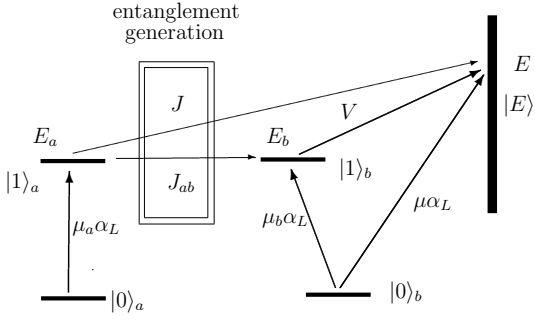}
 \caption{Sketch of an auto-ionization system $ b $ interacting with a
  two-level atom $ a $. Ground states are denoted as $ |0\rangle_a $ and
  $ |0\rangle_b $ whereas symbols $ |1\rangle_a $, $ |1\rangle_b $, and
  $ |E\rangle $ stand for the excited states. Dipole moments $ \mu_a $, $ \mu_b $, and $
  \mu $ describe the appropriate interactions. The excited discrete state at atom $ a $ ($ b $) has
  energy $ E_a $ ($ E_b $), whereas energies $ E $ characterize excited
  states $ |E\rangle $ of the continuum. Symbol $ V $ describes the Coulomb configurational coupling
  between the excited states at atom $ b $. Constants $ J_{ab} $ and $ J $ emerge from
  the dipole-dipole interaction between the atoms $ a $ and $ b $; $ \alpha_L $ is the pumping
  amplitude.}
\label{fig1}
\end{figure}

We consider an atom $ b $ with one auto-ionizing discrete level
that interacts with a neighbor two-level atom $ a $ by the
dipole-dipole interaction (for the scheme, see Fig.~\ref{fig1}).
Both atoms are excited by a stationary optical field. This
composite system can be described by the Hamiltonian $ \hat{H} $,
\begin{equation}    
\hat{H}=\hat{H}_{\rm a-i}+\hat{H}_{\rm t-a}+\hat{H}_{\rm trans}.
\label{1}
\end{equation}
Here, the Hamiltonian $ \hat{H}_{\rm a-i} $ characterizes the
auto-ionization atom:
\begin{eqnarray}    
 \hat{H}_{\rm a-i} &=& E_b|1\rangle_b{}_b\langle1|+ \int
  dE \, E|E\rangle \langle E| \nonumber \\
 & & \hspace{-3mm} \mbox{} + \int dE \,
  \left[V |E\rangle\,{}_b\langle1|+\mbox{H.c.}\right]
  \nonumber \\
 & & \hspace{-3mm} \mbox{} +
  \left[\mu_b \alpha_L \exp(-iE_L t) |1\rangle_b{}_b\langle0|+\mbox{H.c.}\right]
  \nonumber \\
 & & \hspace{-3mm} \mbox{} + \int dE \, \left[ \mu \alpha_L \exp(-iE_L t) |E\rangle {}_b\langle0|+
  \mbox{H.c.}\right] .
\label{2}
\end{eqnarray}
Energy $ E_b $ means the energy difference between the ground
state $ |0\rangle_b $ and the excited discrete state $ |1\rangle_b
$ of atom $ b $. Similarly, energy $ E $ stands for the energy
difference between the state $ |E\rangle $ in the continuum and
the ground state $ |0\rangle_b $. The Coulomb configurational
coupling between the excited states of atom $ b $ is described by
$ V $. The dipole moments between the ground state $ |0\rangle_b $
of atom $ b $ and its excited states are denoted as $ \mu $ and $
\mu_b $. The stationary optical field with its amplitude $
\alpha_L $ oscillates at frequency $ E_L $. We assume $ \hbar = 1
$.

The Hamiltonian $ \hat{H}_{\rm t-a} $ of the neighbor two-level
atom $ a $ introduced in Eq.~(\ref{1}) takes on the form:
\begin{equation}    
 \hat{H}_{\rm t-a} = E_a|1\rangle_a{}_a\langle1|+
  \left[ \mu_a \exp(-iE_Lt) |1\rangle_a{}_a\langle0|+\mbox{H.c.}\right],
\label{3}
\end{equation}
where $ E_a $ means the energy difference between the ground state
$ |0\rangle_a $ and the excited state $ |1\rangle_a $; $ \mu_a $
stands for the dipole moment.

The Hamiltonian $ \hat{H}_{\rm trans} $ in Eq.~(\ref{1})
characterizes the dipole-dipole interaction between electrons at
atoms $ a $ and $ b $:
\begin{eqnarray}   
 \hat{H}_{\rm trans} &=&
  \left(J_{ab}|1\rangle_b{}_b\langle0||0\rangle_a{}_a\langle1|+\mbox{H.c.}
  \right) \nonumber \\
 & & \mbox{}  + \int \, dE
  \left[ J |E\rangle{}_b\langle0|\, |0\rangle_a{}_a\langle1|+
  \mbox{H.c.}\right] .
\label{4}
\end{eqnarray}
In Eq.~(\ref{4}), $ J_{ab} $ ($ J $) quantifies the dipole-dipole
interaction that leads to the excitation from the ground state
$|0\rangle_b $ into the state $ |1\rangle_b $ ($ |E\rangle $) of
atom $ b $ at the cost of the decay of atom $ a $ from the excited
state $ |1\rangle_a $ into the ground state $ |0\rangle_a $.

Following the approach of Ref.~\cite{PerinaJr2011a}, a state
vector $ |\psi\rangle(t) $ of the system at time $ t $ can be
decomposed as
\begin{eqnarray} 
 |\psi\rangle(t) &=& c_{00}(t)|0\rangle_a |0\rangle_b + c_{10}(t)|1\rangle_a
  |0\rangle_b \nonumber \\
 & & \mbox{} + c_{01}(t)|0\rangle_a |1\rangle_b + c_{11}(t)|1\rangle_a |1\rangle_b
  \nonumber \\
 & & \mbox{} + \int \, dE  d_0(E,t)|0\rangle_a |E\rangle
  \nonumber \\
 & & \mbox{} + \int \, dE d_1(E,t)|1\rangle_a |E\rangle
\label{5}
\end{eqnarray}
using time-dependent coefficients $ c_{00} $, $ c_{01} $, $ c_{10}
$, $ c_{11} $, $ d_0(E) $, and $ d_1(E) $.

These coefficients satisfy a system of differential equations
which can be conveniently written in the matrix form:
\begin{eqnarray}  
 \frac{d}{dt} {\bf c}(t) &=& -i{\bf A}{\bf c}(t)-i\int \, dE
  {\bf B} {\bf d}(E,t) ,  \nonumber \\
 \frac{d}{dt} {\bf d}(E,t) &=& -i{\bf B}^\dagger {\bf c}(t)-i{\bf
  K}(E){\bf d}(E,t)
\label{6}
\end{eqnarray}
and
\begin{equation}  
 {\bf c}(t) =\left[\begin{array}{c}
  c_{00}(t)\\c_{10}(t)\\c_{01}(t)\\c_{11}(t)
  \end{array}\right] ,
 \hspace{5mm} {\bf d}(E,t)=\left[\begin{array}{c}
  d_{0}(E,t)\\d_{1}(E,t) \end{array}\right].
\label{7}
\end{equation}
The matrices $ {\bf A} $, $ {\bf B} $, and $ {\bf K} $ introduced
in Eq.~(\ref{6}) are time-independent provided that a basis
rotated at the pump-field frequency $ E_L $ is used:
\begin{eqnarray}   
 & & {\bf A} = \left[\begin{array}{cccc}
  0 & \mu_a^*\alpha_L^* & \mu_b^*\alpha_L^* & 0 \\
  \mu_a\alpha_L & \Delta_a & J_{ab}^* & \mu_b^*\alpha_L^* \\
  \mu_b\alpha_L & J_{ab} & \Delta_b & \mu_a^*\alpha_L^* \\
  0 & \mu_b\alpha_L & \mu_a\alpha_L & \Delta_a + \Delta_b
  \end{array}\right] , \nonumber \\
 & &
\label{8}
  \\
 & & {\bf B} = \left[\begin{array}{cc}
   \mu^*\alpha_L^* & 0 \\ J^* & \mu^*\alpha_L^* \\
   V^* & 0 \\ 0 & V^* \end{array}\right],
\label{9}
   \\
 & & {\bf K}(E) = \left[\begin{array}{cc}
  E-E_L & \mu_a^*\alpha_L^* \\
  \mu_a\alpha_L & E-E_L+\Delta_a \end{array}\right] .
\label{10}
\end{eqnarray}
Here $ \Delta_a = E_a - E_L $ and $ \Delta_b = E_b - E_L $ stand
for the frequency detunings of discrete excited states with
respect to the pump-field frequency.

Contrary to the solution of the model equations found in
\cite{PerinaJr2011a} we adopt here the most general approach based
on algebraic decomposition of dynamical matrices and solution of
the corresponding Sylvester equation. We first neglect threshold
effects in the ionization, eliminate continuum coefficients $ {\bf
d}(E) $ in Eq.~(\ref{6}), and introduce a new matrix $ {\bf M} $:
\begin{equation}  
 {\bf M} = {\bf A} - i\pi{\bf B}{\bf B}^\dagger.
\label{11}
\end{equation}
The matrix $ {\bf M} $ describes the dynamics of only discrete
states that is governed by the vector $ {\bf c} $. We denote
eigenvalues of the matrix ${\bf K}(E) $ as $ E-\xi_1 $ and $
E-\xi_2 $ and eigenvalues of the matrix $ {\bf M} $ as
$\Lambda_{{\bf M}j}$, $j=1,2,3,4$. These eigenvalues occur in the
matrix decompositions of matrices $ {\bf K} $ and $ {\bf M} $:
\begin{eqnarray}  
 {\bf K}(E) &=& (E-\xi_1) {\bf K}_1 + (E-\xi_2) {\bf K}_2,
\label{12} \\
 {\bf M} &=& \sum_{j=1}^4\Lambda_{{\bf M}j}{\bf M}_j.
\label{13}
\end{eqnarray}

The basis matrices $ {\bf K}_1 $ and $ {\bf K}_2 $ can be obtained
from the following equations:
\begin{eqnarray}   
 {\bf K}_1+{\bf K}_2 &=& {\bf I_2}, \nonumber \\
 (E-\xi_1){\bf K}_1+(E-\xi_2){\bf K}_2 &=& {\bf K}(E).
\label{14}
\end{eqnarray}
The eigenvalues $ \xi_1 $ and $ \xi_2 $ are given as follows:
\begin{eqnarray}   
 \xi_{1,2} &=& E_L -\frac{\Delta_a \pm \delta\xi }{2} , \nonumber \\
 \delta \xi &=& \sqrt{ \Delta_a^2 + 4|\mu_a\alpha_L|^2 } .
\label{15}
\end{eqnarray}
Symbol $ \delta\xi $ means the frequency of Rabi oscillations of
the two-level atom $ a $.

Similarly, the basis matrices $ {\bf M}_j $, $ j=1,2,3,4 $, arise
as the solution of the following equations:
\begin{eqnarray}  
 \sum_{j=1}^4{\bf M}_j &=& {\bf I_4}, \nonumber \\
 \sum_{j=1}^4\Lambda_{{\bf M}j}^k{\bf M}_j &=& {\bf M}^k, \hspace{5mm}
  k=1,2,3.
\label{16}
\end{eqnarray}
In Eqs.~(\ref{14}) and (\ref{16}), $ {\bf I_2} $ and $ {\bf I_4} $
are $ 2\times 2 $ and $ 4\times 4 $ unit matrices, respectively.

After the introduction of matrix $ {\bf M} $ in Eq.~(\ref{11}),
the solution of Eqs.~(\ref{6}) for the vector $ {\bf c} $ can be
written in the very simple form:
\begin{equation}  
 {\bf c}(t) = \exp(-i{\bf M}t){\bf c}(0);
\label{17}
\end{equation}
$ {\bf c}(0) $ is the vector of initial conditions.

On the other hand, a newly introduced matrix $ {\bf T}(E) $ (of
dimension $ 2\times 4 $) obtained as the solution to the Sylvester
equation \cite{Gantmacher2000}
\begin{equation}  
 {\bf K}(E){\bf T}(E) - {\bf T}(E){\bf M} = {\bf B}^\dagger
\label{18}
\end{equation}
is useful for expressing the solution of Eqs.~(\ref{6}) for the
continuum of states described by the vector $ {\bf d}(E) $. On
using the matrix decompositions written in Eqs.~(\ref{12}) and
(\ref{13}), the solution of the Sylvester equation (\ref{18}) can
be expressed as follows:
\begin{equation} 
 {\bf T}(E) = \sum_{k=1}^2\sum_{j=1}^4
  \frac{1}{E - \xi_k - \Lambda_{{\bf M}j}}{\bf K}_k{\bf B}^\dagger{\bf
  M}_j.
\label{19}
\end{equation}
The components of amplitude spectrum of an ionized electron at
atom $ b $ are given by the coefficients in the vector $ {\bf
d}(E,t) $. They can be written in the most general form
\begin{equation} 
 {\bf d}(E,t)=\Bigl( \exp[-i{\bf K}(E)t]{\bf
  T}(E) - {\bf T}(E)\exp[-i{\bf M}t] \Bigr)
  {\bf c}(0)
\label{20}
\end{equation}
depending on the initial conditions. We have assumed that $
d_0(E,0) = d_1(E,0)=0 $ in the derivation of Eq.~(\ref{20}).

As the interaction processes between the discrete states and the
continuum of states are irreversible, the eigenvalues of matrix $
{\bf M} $ are complex with negative imaginary parts. As a
consequence, the expression in Eq.~(\ref{20}) for the amplitude
spectral components $ {\bf d} $ simplifies in the long-time limit:
\begin{equation} 
{\bf d^{lt}}(E,t) = \exp[-i{\bf K}(E)t]{\bf T}(E){\bf c}(0).
\label{21}
\end{equation}

\section{Negativity of a bipartite system in
discrete and continuous Hilbert spaces}

We need to quantify the amount of entanglement between the
two-level atom $ a $ and the auto-ionization atom $ b $ that has a
continuous spectrum. The philosophy based on declinations of the
partially-transposed statistical operators of entangled states
from the positive-semidefinite partially-transposed statistical
operators of separable states \cite{Hill1997,Vidal2002} has been
found fruitful here and has resulted in the definition of
negativity.

Following the approach by Hill and Wooters \cite{Hill1997}, we
write a matrix $ {\bf P} $ of the statistical operator describing
an electron at atom $ a $ and a (fully) ionized electron at atom $
b $ in a given time $ T $ [$ d_j(E) \equiv d_j(E,T) $, $ j=0,1 $]:
\begin{equation}    
 {\bf P} = \left[\begin{array}{cc} d_0(E)d_0^*(E') & d_0(E)d_1^*(E') \\
  d_1(E)d_0^*(E') & d_1(E)d_1^*(E') \end{array}\right].
\label{22}
\end{equation}
We note that the frequencies $ E $ and $ E' $ in Eq.~(\ref{22})
are considered as continuous indices of the matrix $ {\bf P} $.

A partially-transposed matrix $ {\bf P^{Ta}} $ transposed with
respect to the indices of two-level atom $ a $ is obtained after
the exchange of sub-matrices in the upper-left and lower-right
corners of the matrix $ {\bf P} $ in Eq.~(\ref{22}):
\begin{equation}    
 {\bf P^{Ta}} = \left[\begin{array}{cc} d_0(E)d_0^*(E') & d_1(E)d_0^*(E') \\
  d_0(E)d_1^*(E') & d_1(E)d_1^*(E') \end{array}\right].
\label{23}
\end{equation}

In order to determine negativity $ N $, we need to find the
eigenvalues $ \lambda $ of matrix $ {\bf P^{Ta}} $ first. An
eigenvalue $ \lambda $ together with its eigenvector $
(u_0(E),u_1(E)) $ fulfil the following system of equations with a
continuous index $ E $:
\begin{eqnarray}   
 d_0(E)  \int \, dE' d_0^*(E')u_0(E') \hspace{2cm} & &
  \nonumber \\
 \mbox{} + d_1(E)\int \, dE' d_0^*(E')u_1(E') &=& \lambda u_0(E),
  \nonumber \\
 d_0(E) \int \, dE' d_1^*(E')u_0(E') \hspace{2cm} & & \nonumber \\
 \mbox{} + d_1(E)\int \, dE' d_1^*(E')u_1(E') &=& \lambda u_1(E).
  \nonumber \\
 & &
\label{24}
\end{eqnarray}
Integrals in Eqs.~(\ref{24}) give the coefficients $ a_{jk} $ of
the decomposition of eigenvector functions $ u_j(E) $ in the basis
$ d_j(E) $:
\begin{eqnarray}   
 a_{jk} = \int \, dE' d_j^*(E')u_k(E'), \hspace{5mm} j,k=0,1.
\label{25}
\end{eqnarray}
Using the coefficients $ a_{jk} $ defined in Eq.~(\ref{25}), the
equations in (\ref{24}) can be rewritten as follows:
\begin{eqnarray}  
 d_0(E)a_{00} + d_1(E) a_{01} &=& \lambda u_0(E),
  \nonumber \\
 d_0(E)a_{10} + d_1(E)a_{11} &=& \lambda u_1(E).
\label{26}
\end{eqnarray}
The projection of equations in Eq.~(\ref{26}) onto the basis
vectors $ d_j(E) $ results in a system of four algebraic equations
for the coefficients $ a_{jk} $ determining the eigenvector $
(u_0(E),u_1(E)) $:
\begin{equation} 
 \left[\begin{array}{cccc}
  b_{00}&b_{01}&0&0\\
  0&0&b_{00}&b_{01}\\
  b_{10}&b_{11}&0&0\\
  0&0&b_{10}&b_{11}
  \end{array}\right] \left[\begin{array}{c} a_{00}\\a_{01}\\a_{10}\\a_{11}
   \end{array}\right] = \lambda
  \left[\begin{array}{c} a_{00}\\a_{01}\\a_{10}\\a_{11} \end{array}\right].
\label{27}
\end{equation}
The coefficients $ b_{jk} $ introduced in Eq.~(\ref{27}) are the
overlap integrals between the functions $ d_0(E) $ and $ d_1(E) $:
\begin{equation}  
 b_{jk}=\int \, dE d_j^*(E)d_k(E) .
\label{28}
\end{equation}
It holds that $ b_{01} = b_{10}^* $ and $ b_{00}+b_{11} = 1 $ due
to the normalization.

The system of algebraic equations (\ref{27}) has a nontrivial
solution provided that the eigenvalues $ \lambda $ are solutions
of the secular equation:
\begin{equation}   
 \lambda^4- \lambda^3+ {\cal D} \lambda- {\cal D}^2 = 0,
\label{29}
\end{equation}
where
\begin{equation}  
 {\cal D} = b_{00}b_{11}-b_{01}b_{10} .
\label{30}
\end{equation}
The fourth-order polynomial in Eq.~(\ref{29}) can be written as a
product of the second-order polynomials $ (\lambda^2 - {\cal D})
(\lambda^2 - \lambda + {\cal D}) $. This allows to find its roots:
\begin{eqnarray}  
 \lambda_{1,2} &=& \pm \sqrt{\cal D} \nonumber , \\
 \lambda_{3,4} &=& \frac{1}{2} \pm \sqrt{\frac{1}{4} - {\cal D}} .
\label{31}
\end{eqnarray}
As the negativity $ N $ is given by the amount of negativeness in
the eigenvalues $ \lambda $, we have
\begin{equation}   
 N = \sqrt{\cal D} .
\label{32}
\end{equation}
Alternative and more intuitive derivation of the formula in
Eq.~(\ref{32}) can be found in Appendix~A invoking the
decomposition of functions $ d_0(E) $ and $ d_1(E) $.

In parallel to the entanglement, quantum discord
\cite{Ollivier2001} has been discussed in the last years for
systems composed of several parts \cite{AlQasimi2011}. Discord
quantifies the amount of information in the whole system that
cannot be extracted using quantum measurements at separated parts.
Provided that a bipartite system is in a pure state quantum
discord is quantified by entropy $ S $ of entanglement. The
entropy $ S $ of entanglement is given by the entropy of reduced
statistical operator $ \varrho^a $ of atom $ a $ that takes the
form
\begin{equation}   
 \varrho^a = \left[ \begin{array}{cc} b_{00} & b_{10} \\
  b_{01} & b_{11} \end{array} \right]
\label{33}
\end{equation}
exploiting the coefficients $ b_{jk} $. The eigenvalues $
\lambda_{3,4} $ written in Eq.(\ref{31}) naturally give also the
eigenvalues of matrix $ \varrho^a $ and so they can be
conveniently used in expressing the entropy $ S $. The entropy $ S
$ of entanglement is given by the usual formula $ S = -
\sum_{j=3,4} \lambda_j \log_2(\lambda_j) $, $ \log_2 $ being the
logarithm of base two. This formula provides us the following
expression:
\begin{eqnarray}    
 S &=& - \frac{1}{2} \left[ \log_2({\cal D}) + \sqrt{1-4{\cal D}}
  \log_2 \left( \frac{ 1+\sqrt{1-4{\cal D}} }{ 1- \sqrt{1-4{\cal D}}
  } \right) \right] . \nonumber \\
 & &
\label{34}
\end{eqnarray}
Here, determinant $ {\cal D} $ of the matrix $ \varrho^a $ is
given in Eq.~(\ref{30}).

\begin{figure}  
 \includegraphics[scale=0.3]{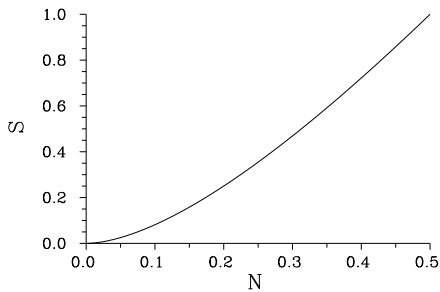}
 \caption{Entropy $ S $ of entanglement as a function of negativity $ N $
  in the interval $ [0,0.5] $ of attainable values of $ N $.}
\label{fig2}
\end{figure}

Combining Eqs.~(\ref{32}) and (\ref{34}), the entropy $ S $ of
entanglement can be expressed as a monotonous function of
negativity $ N $ (see Fig.~\ref{fig2}):
\begin{eqnarray}    
 S &=& - \log_2(N) - \frac{ \sqrt{1-4N^2} }{2}
  \log_2\left( \frac{ 1+\sqrt{1-4N^2} }{ 1- \sqrt{1-4N^2}
  } \right) .  \nonumber  \\
 & &
\label{35}
\end{eqnarray}
The curve in Fig.~\ref{fig2} reveals that both quantities can be
equally well used for the quantification of entanglement in the
considered system.

The negativity $ N $ can also be expressed in terms of eigenvalues
of the Schmidt decomposition of the state $ |\psi\rangle $ in the
long-time limit. Substituting Eq.~(\ref{30}) into Eq.~(\ref{32}),
we arrive at the useful formula for negativity $ N $:
\begin{equation}   
 N = \sqrt{ \frac{1}{2} \sum_{j,k=0}^{1} \left[ b_{jj}b_{kk}
  - b_{jk}b_{kj} \right] }.
\label{36}
\end{equation}
Further substitution for the coefficients $ b_{jk} $ from
Eq.~(\ref{28}) provides the negativity $ N $ depending on the
reduced statistical operator $ \varrho^b $ of the continuum:
\begin{eqnarray}   
 N &=& \sqrt{ \frac{1}{2} \left[ 1
  - \int dE \, \int dE' \, |\varrho^b(E,E')|^2 \right] },\,\,\,
\label{37}  \\
 \varrho^b(E,E') &=& \sum_{k=0,1} d_k(E) d_k^*(E') .
\label{38}
\end{eqnarray}
Using the coefficients $ \sqrt{\lambda_3} $ and $ \sqrt{\lambda_4}
$ of the Schmidt decomposition of the state $ |\psi\rangle $, the
formula (\ref{37}) can be recast into the simple form:
\begin{equation}  
 N = \sqrt{\lambda_3 \lambda_4} .
\label{39}
\end{equation}

The formula for negativity $ N $ in Eq.~(\ref{32}) can even be
used for finite times $ t $, in which discrete states of atom $ b
$ are populated. In this case, the formula in Eq.~(\ref{28}) has
to be replaced by the more general one:
\begin{equation} 
 b_{jk}= \sum_{l} c^*_{jl} c_{kl} + \int \, dE d_j^*(E)d_k(E) .
\label{40}
\end{equation}

\section{Quadratic negativity and its spectral density}

The substitution of expression in Eq.~(\ref{30}) into the formula
(\ref{32}) for negativity $ N $ gives us an expression that
indicates the existence of quadratic negativity $ N_q $ as a
measure of entanglement that allows to introduce a spectral
density \cite{Dur1999}:
\begin{equation}   
 N_q \equiv 4N^2 = 4(b_{00} b_{11} - b_{01} b_{10} )  .
\label{41}
\end{equation}
The use of expressions~(\ref{28}) for the coefficients $ b_{jk} $
allows us to rewrite the formula in Eq.(\ref{41}) as:
\begin{equation}   
 N_q = 2 \int dE\, \varrho(E) \int \, dE' \varrho(E')
   n_q(E,E') ,
\label{42}
\end{equation}
where $ \varrho(E) $ gives the density of states $ |E\rangle $ in
the continuum:
\begin{equation}    
 \varrho(E) = \sum_{j=0}^{1} |d_j(E)|^2 .
\label{43}
\end{equation}
The spectral density $ n_q(E,E') $ of quadratic negativity
introduced in Eq.~(\ref{42}) is obtained in the form:
\begin{eqnarray}  
 n_q(E,E') &=& \frac{1}{\varrho(E)\varrho(E')} \left[
  \sum_{j,k=0}^{1} |d_j(E)|^2 |d_k(E')|^2 \right.
  \nonumber \\
 & & \hspace{-1cm} \left.  \mbox{} - \sum_{j,k=0}^{1} d^*_j(E)d_k(E) d^*_k(E')d_j(E')
  \right] .
\label{44}
\end{eqnarray}
The value of spectral density $ n_q(E,E') $ of quadratic
negativity gives the value of quadratic negativity of a
qubit-qubit system composed of the states $ \{ |0\rangle_a,
|1\rangle_a \} $ and $ \{ |E\rangle , |E'\rangle \} $. According
to Eq.~(\ref{42}), the quadratic negativity $ N_q $ is given as a
weighted sum of quadratic qubit-qubit negativities between the
two-level atom $ a $ and all possible qubits embedded inside the
continuum. This interpretation is important from the physical
point of view, because it allows to interpret the overall
entanglement as composed of individual spectral contributions. We
note that values of both the quadratic negativity $ N_q $ and its
density $ n_q $ lie in the interval $ [0,1] $. We also note that
an alternative normalization in the definition (\ref{44}) of
density $ n_q $ of quadratic negativity is possible. It is based
on substituting the factor $ 1/[\varrho(E)\varrho(E')] $ by the
factor $ 4/[\varrho(E)+\varrho(E')]^2 $. However, this
'mathematically more compact' normalization is not suitable for
indicating entanglement in the case of qubits with considerably
different values of the probability densities $ \varrho(E) $ and $
\varrho(E') $.

Experimental determination of the density $ n_q $ of quadratic
negativity has to take into account a finite resolution $ \Delta E
$ of frequencies of free electrons. That is why, it is convenient
to introduce a series of experimental quadratic negativities $
N_q^{(i)} $, $ i=1,2,\ldots $, that are obtained after spectral
filtering of a free electron by using $ i $ filters positioned at
the central frequencies $ E_k $, $ k=1,\ldots, i $:
\begin{eqnarray}   
 N^{(i)}_q(E_1,\ldots, E_i) &=&  \nonumber \\
 & & \hspace{-3.5cm} \Bigl[ b_{00}^{(i)}(E_1,\ldots,E_i)
   b_{11}^{(i)}(E_1,\ldots,E_i) - |b_{01}^{(i)}(E_1,\ldots,E_i)|^2 \Bigr]^{1/2}
   \nonumber \\
 & & \hspace{-3.5cm} \mbox{} \times \Bigl[ b_{00}^{(i)}(E_1,\ldots,E_i) +
     b_{11}^{(i)}(E_1,\ldots,E_i) \Bigr]^{-1} .
\label{45}
\end{eqnarray}
The coefficients $ b_{jk}^{(i)}(E_1,\ldots,E_i) $ occurring in
Eq.~(\ref{45}) depend on the experimental frequency width $ \Delta
E $ and are given as:
\begin{equation}  
 b_{jk}^{(i)}(E_1,\ldots,E_i) = \sum_{l=1}^{i} \int_{E_l-\Delta E/2}^{E_l+\Delta E/2} \,
  dE' d_j^*(E')d_k(E').
\label{46}
\end{equation}
We note that the last term in the expression (\ref{45}) originates
in the normalization of the considered state.

\section{Entanglement generation}

The entanglement between electrons at atoms $ a $ and $ b $ is
generated by the dipole-dipole interaction that is characterized
by the coefficients $ J_{ab} $ and $ J $. This means that two
different channels of the entanglement generation exist. In the
first channel, the entanglement among the discrete states at atoms
$ a $ and $ b $ is formed due to the dipole-dipole interaction
described by the coefficient $ J_{ab} $ first. Subsequently, this
entanglement is transferred to the continuum of states $ |E\rangle
$ using either the Coulomb interaction ($ V $) or the optical
dipole interaction ($ \mu\alpha_L $). The second channel is based
on the dipole-dipole interaction ($ J $) between the excited
discrete state $ |1\rangle_a $ at atom $ a $ and the continuum of
states $ |E\rangle $ at the ionization atom $ b $.

The dynamics of the system is such that an electron at atom $ b $
gradually 'leaks' into the continuum of states $ |E\rangle $. The
probability of finding this electron in a combination of discrete
states $ |0\rangle_b $ and $ |1\rangle_b $ decreases roughly
exponentially. After a sufficiently long time, this probability is
practically zero, the electron is fully ionized and its long-time
spectrum completely characterizes its state. On the other hand,
the electron at atom $ b $ periodically oscillates between its
discrete states in a stationary optical field at the Rabi
frequency. The entanglement between the bound electron at atom $ a
$ and the ionized electron at atom $ b $ is formed during the
period of ionization and is 'frozen' as soon as atom $ b $ is
completely ionized. At this instant, the entanglement reaches its
long-time limit, but superimposed periodic oscillations are
possible under certain conditions (see below).

\begin{figure}  
 (a) \includegraphics[scale=0.3]{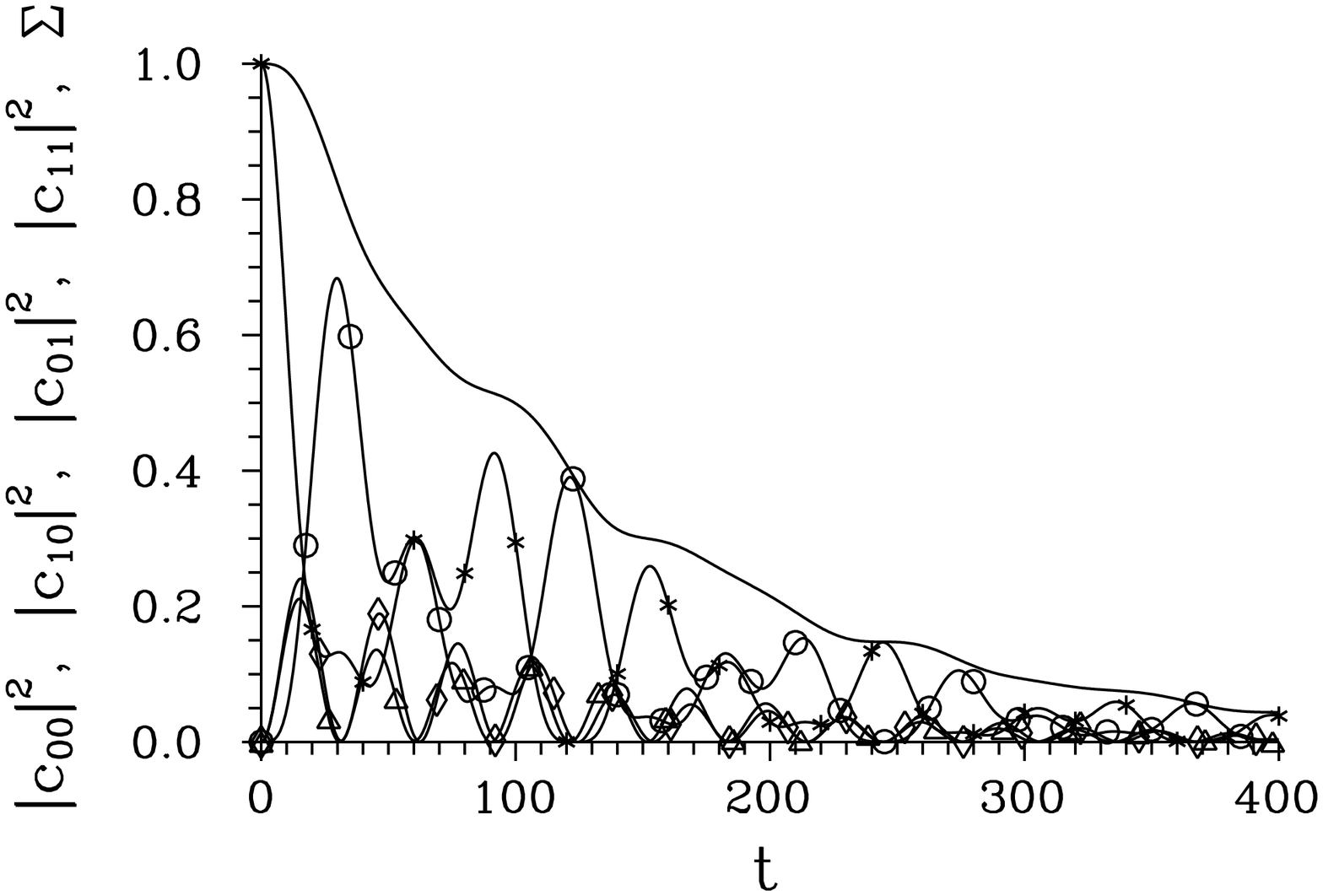}

 \vspace{7mm}
 (b) \includegraphics[scale=0.3]{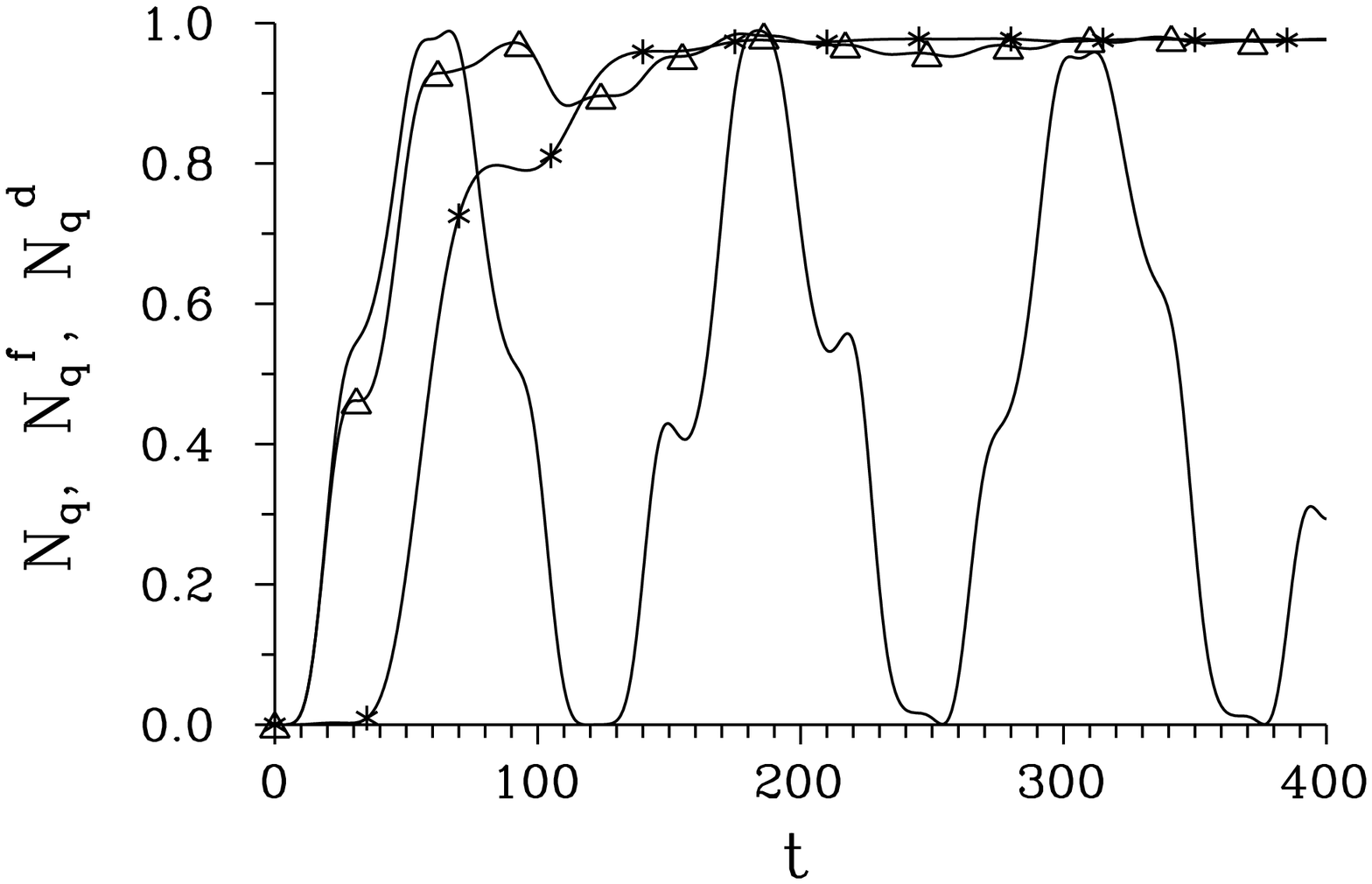}

 \caption{Temporal evolution of (a) probabilities $ |c_{00}|^2 $ (solid curve
  with $ \ast $), $ |c_{10}|^2 $ (solid curve with $ \diamond $),
  $ |c_{01}|^2 $ (solid curve
  with $ \triangle $), $ |c_{11}|^2 $ (solid curve with $ \circ $)
  of detecting two electrons
  in the appropriate discrete states and their sum $ \Sigma $
  ($ \Sigma = \sum_{j=0,1} \sum_{k=0,1} |c_{jk}|^2 $, solid
  curve) and (b) quadratic negativities $ N_q $ (solid curve
  with $ \ast $), $ N_q^f $ (solid curve
  with $ \triangle $), and $ N_q^d $ (solid curve); $ \mu_a\alpha_L =
  \mu_b\alpha_L = J_{ab} = V = 0.05 $,
  $ \mu = J = 0 $, $ E_a = E_b = E_L = 1 $.}
\label{fig3}
\end{figure}

Let us concentrate on the first channel. Both electrons at atoms $
a $ and $ b $ being initially in their ground states gradually
move into their excited states $ |1\rangle_a |0\rangle_b $, $
|0\rangle_a |1\rangle_b $, and $ |1\rangle_a |1\rangle_b $ due to
the interaction with the stationary optical field [see
Fig.~\ref{fig3}(a)]. The entanglement between discrete states
arises from the dipole-dipole interaction between the states $
|1\rangle_a |0\rangle_b $ and $ |0\rangle_a |1\rangle_b $. The
probabilities $ |c_{10}|^2 $ and $ |c_{01}|^2 $ affiliated to
these states periodically return to zero with a period that
decreases with the increasing values of $ |J_{ab}| $, $
|\mu_a\alpha_L| $, and $ |\mu_b\alpha_L| $. At these instants,
highly entangled states occur and their quadratic negativities $
N_q^d $ quantifying entanglement among discrete states reach local
maxima [see Fig.~\ref{fig3}(b)]. Provided that the probabilities
of the ground state $ |0\rangle_a |0\rangle_b $ and the state with
both electrons excited are balanced ($ |c_{00}|^2 \approx
|c_{11}|^2 $), the quadratic negativity $ N_q^d $ reaches its
maximum value one. The quadratic negativity $ N_q^d $ oscillates
between its maximum and zero during the time evolution. The
entanglement between the discrete states at atom $ a $ and the
continuum of states at atom $ b $ arises as a consequence of the
interaction of the continuum of states $ |E\rangle $ with the
discrete states $ |0\rangle_b $ and $ |1\rangle_b $. The quadratic
negativity $ N_q $ appropriate for this entanglement typically
increases during the time evolution and gradually reaches its
long-time value, as documented in Fig.~\ref{fig3}(b). However,
weak oscillations may occur in this evolution. The overall
quadratic negativity $ N_q^f $ that characterizes the entanglement
between atoms $ a $ and $ b $ including all states, behaves
similarly as the quadratic negativity $ N_q $ comprising only the
continuum of states. As a rule of thumb, a slightly stronger
optical pumping of atom $ a $ compared to atom $ b $ ($ \mu_a >
\mu_b $) results in greater values of the long-time quadratic
negativity $ N_q^{\rm lt} $.

\begin{figure}  
 \includegraphics[scale=0.3]{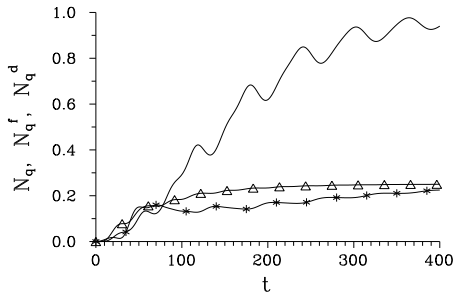}
 \caption{Temporal evolution of quadratic negativities $ N_q $ (solid curve
  with $ \ast $), $ N_q^f $ (solid curve
  with $ \triangle $), and $ N_q^d $ (solid curve); $ J = 0.05 $, $ J_{ab} = 0 $,
  values of the other parameters are the same as in the caption to Fig.~\ref{fig3}.}
\label{fig4}
\end{figure}

In the second channel, the entanglement is generated directly by
the dipole-dipole interaction between the excited state $
|1\rangle_a $ and the continuum of states $ |E\rangle $. The
ability to generate the entanglement is weaker compared to the
first channel. 'Transfer of entanglement' can be observed also
here and so nonzero values of the quadratic negativity $ N_q^d $
are found during the temporal evolution [see Fig.~\ref{fig4}].
Even the maximum entangled discrete states ($ N_q^d = 1 $) can be
reached. This clearly shows that there exists a strong
'back-action' from the 'reservoir' continuum of states $ |E\rangle
$ towards the discrete states $ |0\rangle_b $ and $ |1\rangle_b $.
Otherwise, the observed temporal evolution is qualitatively
similar to that found in the first channel.

Some general features of the behavior of quadratic negativity $
N_q $ in the long-time limit can be obtained even analytically. A
detailed analysis of the long-time solution in Eq.~(\ref{21}) has
shown \cite{PerinaJr2011a} that the coefficients $ b_{00} $ and $
b_{11} $ giving the probabilities of finding an electron at atom $
a $ in the states $ |0\rangle_a $ and $ |1\rangle_a $,
respectively, can be expressed in the form:
\begin{eqnarray}   
 b_{00}^{\rm lt}(t) &=& a + \left[ b\exp(i\delta\xi t) + \mbox{c.c.} \right],
  \nonumber \\
 b_{11}^{\rm lt}(t) &=& (1-a) - \left[ b\exp(i\delta\xi t) + \mbox{c.c.}
 \right].
\label{47}
\end{eqnarray}
Constant $ a $ describes the steady-state parts of probabilities $
b_{00} $ and $ b_{11} $, whereas constant $ b $ gives the amount
of probability that oscillates between the states $ |0\rangle_a $
and $ |1\rangle_a $ at the Rabi frequency $ \delta\xi $. The
symbol $ {\rm c.c.} $ replaces a complex conjugate term. On the
other hand, the cross-correlation coefficient $ b_{01} $ can be
written as
\begin{equation}   
 b_{01}(t) = c_1 + c_2 \exp(i\delta\xi t) + c_3 \exp(-i\delta\xi t)
 ;
\label{48}
\end{equation}
$ c_1 $, $ c_2 $, and $ c_3 $ being constants.

Using the equality $ b^2 = - c_2c_3^* $ valid in the model, we
arrive at the following formula for the long-time quadratic
negativity $ N_q^{\rm lt} $:
\begin{eqnarray}   
 N_q^{\rm lt}(t) &=& 4 \left\{ a(1-a) - 2|b|^2 - |c_1|^2 - |c_2|^2 -
 \frac{|b|^4}{|c_2|^2}  \right. \nonumber \\
 & & \hspace{-1cm} \left. \mbox{} + \left[ (1-2a)b-c_1^*c_2
  + \frac{c_1b^2}{c_2} \right] \exp(i\delta\xi t)
  \right\} .
\label{49}
\end{eqnarray}
We can see from Eq.~(\ref{49}) that the quadratic negativity $
N_q^{\rm lt} $ is composed of a steady-state part and an
oscillating part with the Rabi frequency $ \delta\xi $. However,
the oscillating part is usually much smaller than the steady-state
one. Even if atom $ a $ is resonantly pumped, the oscillating term
in Eq.~(\ref{49}) vanishes and we arrive at the simplified
formula:
\begin{equation}   
 N_q^{\rm lt,res} = 4 \left[ a(1-a) - 2|b|^2 - |c_1|^2 - |c_2|^2 -
  \frac{|b|^4}{|c_2|^2} \right].
\label{50}
\end{equation}
According to Eq.~(\ref{50}), equal steady-state probabilities $ a
$ and $ (1-a) $ of detecting the electron at atom $ a $ in the
states $ |0\rangle_a $ and $ |1\rangle_a $, respectively, are
needed to reach the maximum value of quadratic negativity $
N_q^{\rm lt} $ ($ a=1/2 $). Moreover, nonzero values of constants
$ |b| $, $ |c_1| $ and $ |c_2| $ lower the values of long-time
quadratic negativity $ N_q^{\rm lt} $.

\begin{figure}  
 (a) \includegraphics[scale=0.25]{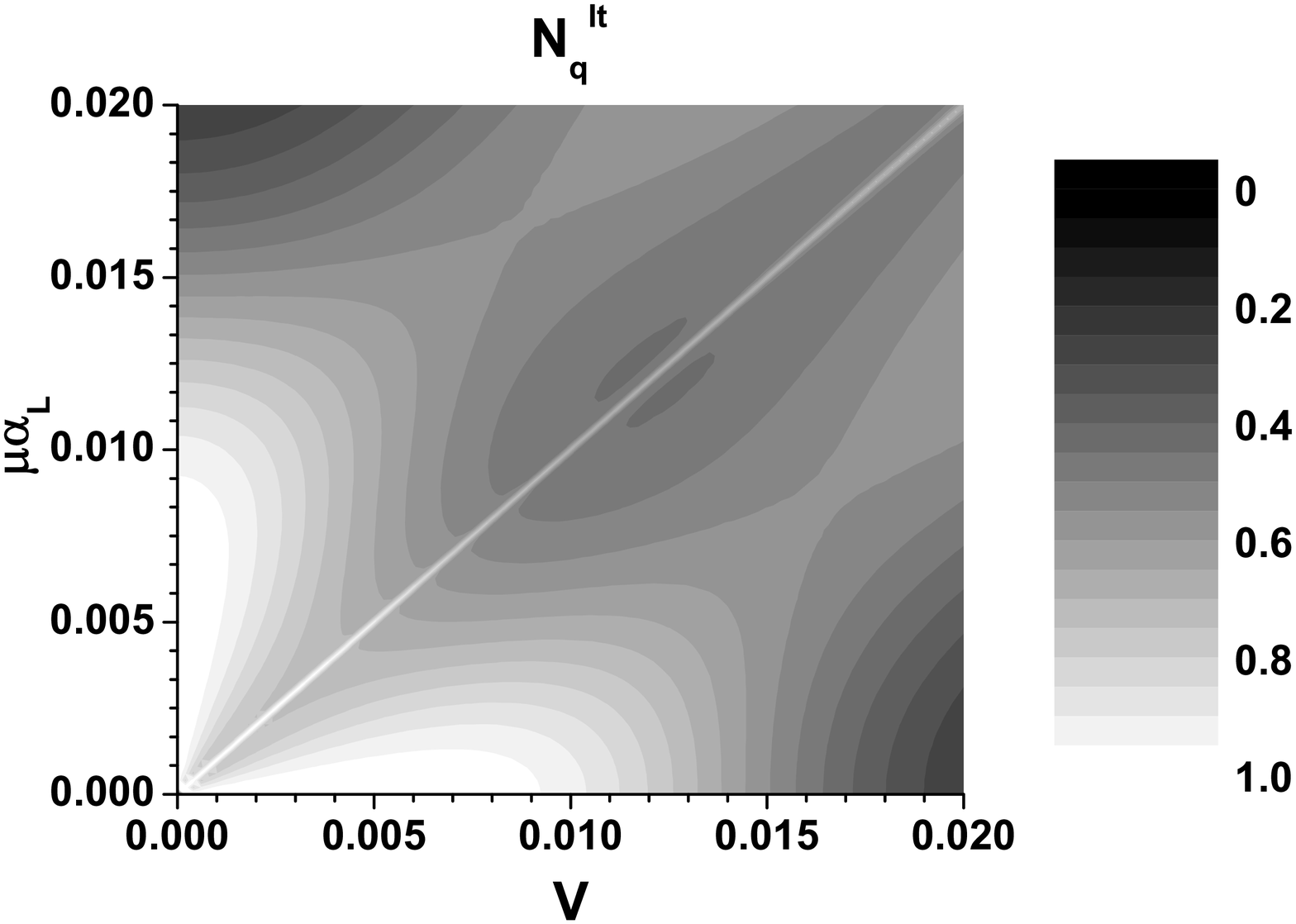}

 \vspace{7mm}
 (b) \includegraphics[scale=0.25]{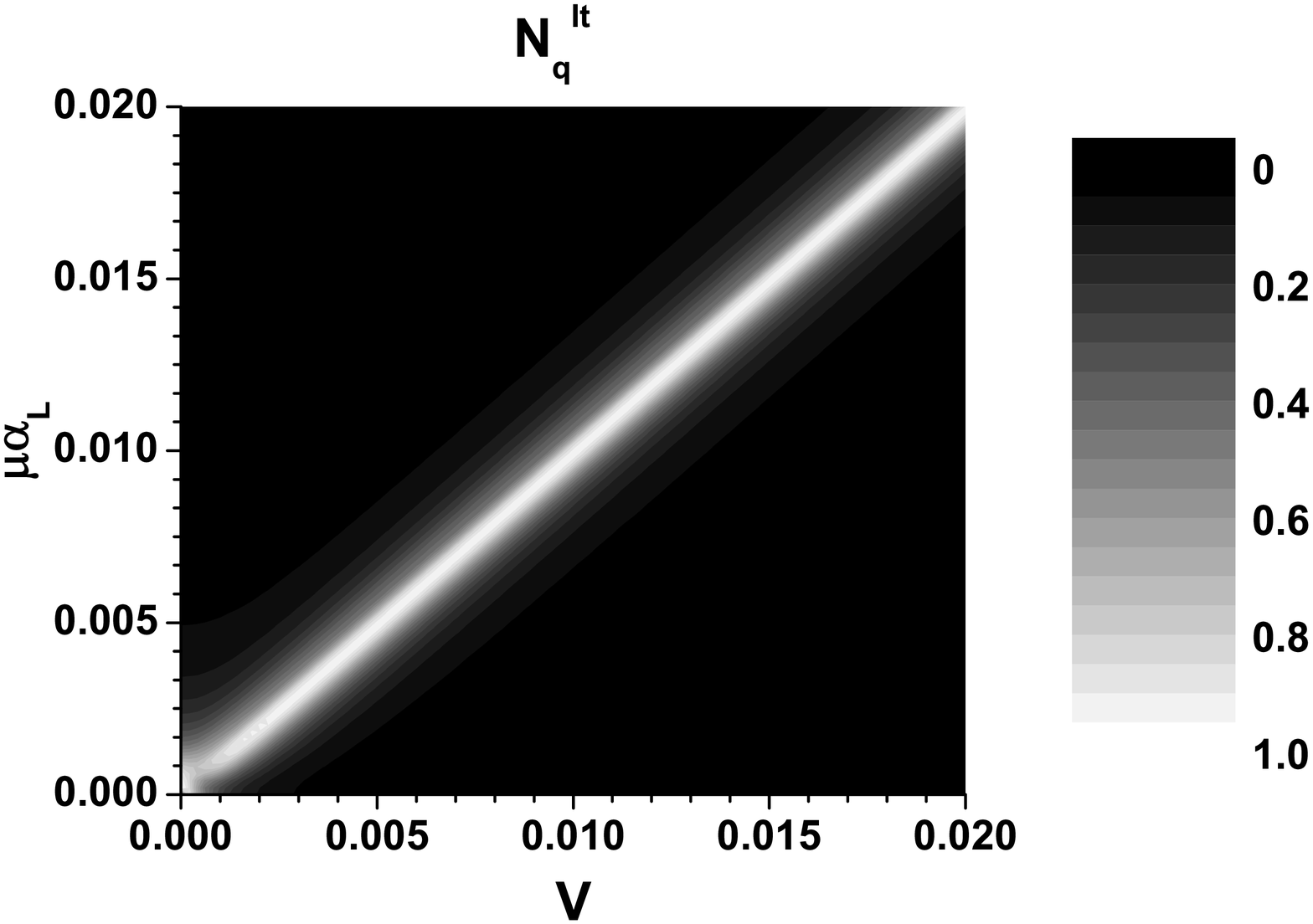}

 \vspace{7mm}
 (c) \includegraphics[scale=0.25]{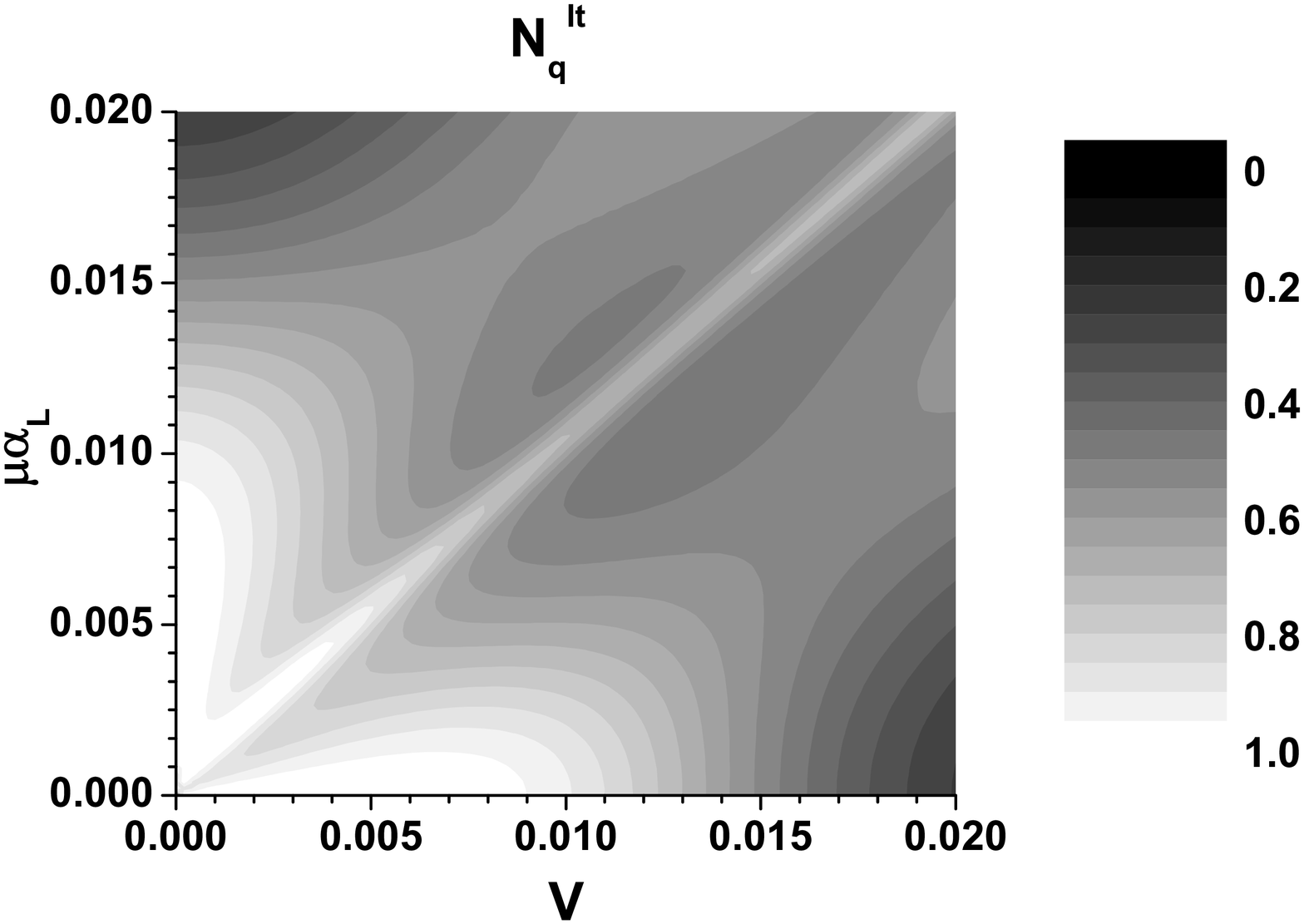}
 \caption{Topo graphs of long-time quadratic negativity $ N_q^{\rm lt} $ depending
  on optical pumping $ \mu\alpha_L $ and strength $ V $ of the Coulomb interaction
  for (a) $ J_{ab} = 0.001 $, $ J = 0 $, (b) $ J_{ab} = 0 $,
  $ J = 0.001 $, and (c) $ J_{ab} = J = 0.001 $; $ \mu_a\alpha_L =
  \mu_b\alpha_L = 0.05 $, $ E_a = E_b = E_L = 1 $.}
\label{fig5}
\end{figure}

The numerical analysis of the long-time behavior of quadratic
negativity $ N_q^{\rm lt} $ has revealed that the larger the
values of dipole-dipole constants $ J_{ab} $ and $ J $ are, the
larger is the potential to generate highly entangled states. In
order to arrive at high values of the quadratic negativity $
N_q^{\rm lt} $, the values of constants $ \mu\alpha_L $ and $ V $
have to be sufficiently small compared to the values of $ J $ and
$ J_{ab} $. This can be physically explained as follows. The
constants $ \mu\alpha_L $ and $ V $ determine the speed of
transfer of an electron at atom $ b $ into the continuum of states
$ |E\rangle $. If this speed is too fast, the electron at atom $ b
$ has not enough time to create the entanglement with the electron
at atom $ a $. As a consequence, the entanglement between two
electrons is weaker. This behavior is documented in
Fig.~\ref{fig5} considering both channels of entanglement
generation. However, the graphs in Fig.~\ref{fig5} reveal that
also greater values of the constants $ \mu\alpha_L $ and $ V $
allow to reach strong entanglement under the condition $
\mu\alpha_L \approx V $. The analysis of temporal behavior of the
system has shown that the movement of the electron at atom $ b $
into the continuum of states $ |E\rangle $ is considerably slowed
down in this case of balanced interactions $ \mu\alpha_L $ and $ V
$. This slowing-down then gives enough time for the entanglement
generation even for smaller values of the constants $ J_{ab} $ and
$ J $. This regime is even preferred for the channel exploiting
the constant $ J $, as the graph in Fig.~\ref{fig5}(b) shows.

\begin{figure}  
 \includegraphics[scale=0.25]{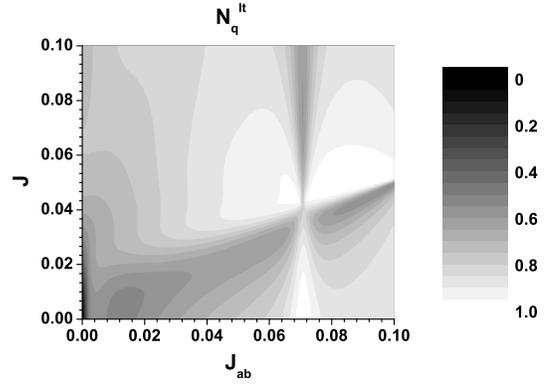}
 \caption{Topo graphs of long-time quadratic negativity $ N_q^{\rm lt} $ as it depends
  on dipole-dipole coupling constants $ J_{ab} $ and $ J $; $ \mu_a\alpha_L =
  \mu_b\alpha_L = \mu\alpha_L = 0.05 $, $ V=0.01 $,  $ E_a = E_b = E_L = 1 $.}
\label{fig6}
\end{figure}

Two channels based on the constants $ J_{ab} $ and $ J $ mutually
'interfere' in creating the entanglement between two electrons.
This can be conveniently used for reaching greater values of the
quadratic negativity $ N_q^{\rm lt} $ in regions, where the above
described conditions are not met. Great values of the quadratic
negativity $ N_q^{\rm lt} $ can be obtained in specific areas of
the space spanned by the constants $ J_{ab} $ and $ J $, as
illustrated in Fig.~\ref{fig6}.

\begin{figure}  
 \includegraphics[scale=0.25]{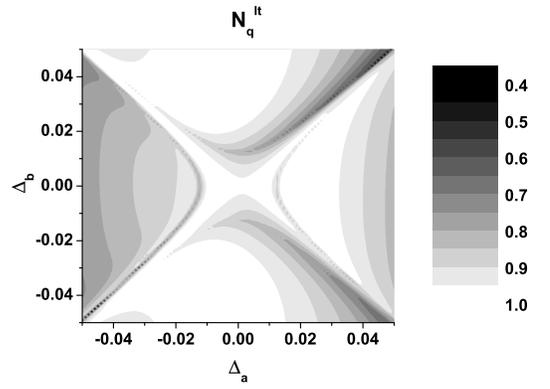}
 \caption{Topo graph of long-time quadratic negativity $ N_q^{\rm lt} $ as a function of
  detunings $ \Delta_a $ and $ \Delta_b $ of atoms $ a $ and $ b $, respectively;
  $ \mu_a\alpha_L = \mu_b\alpha_L = 0.05 $, $ \mu\alpha_L = 0.005 $,
  $ J_{ab} = V=0.001 $, $ J = 0 $, $ E_L = 1 $.}
\label{fig7}
\end{figure}

We have considered the resonant pumping of atoms $ a $ and $ b $
up to now. The non-resonant pumping of both atoms makes the
dynamics as well as the entanglement generation even more complex.
Upon depending on conditions, the frequency detuning of atoms $ a
$ and $ b $ may either support the entanglement creation or
degrade it. A typical graph showing the behavior of quadratic
negativity $ N_q^{\rm lt} $ in dependence on the detunings $
\Delta_a $ and $ \Delta_b $ is plotted in Fig.~\ref{fig7}.

\section{Spectral entanglement}

\begin{figure}  
 \includegraphics[scale=0.25]{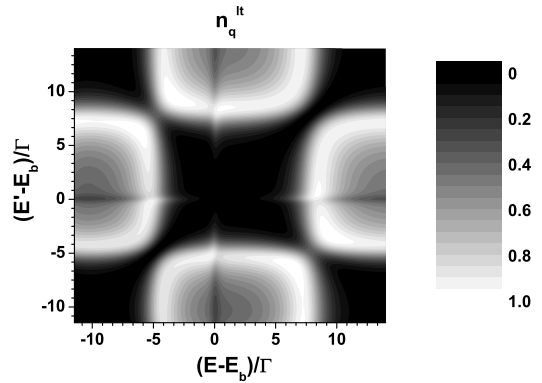}

 \caption{Topo graph of density $ n^{\rm lt}_q $ of quadratic negativity
  showing qubit-qubit correlations in relative frequencies $ (E-E_b)/\Gamma $
  and $ (E'-E_b)/\Gamma $, values of parameters given in the
  caption to Fig.~\ref{fig3} are used.}
\label{fig8}
\end{figure}

\begin{figure}  
 \includegraphics[scale=0.25]{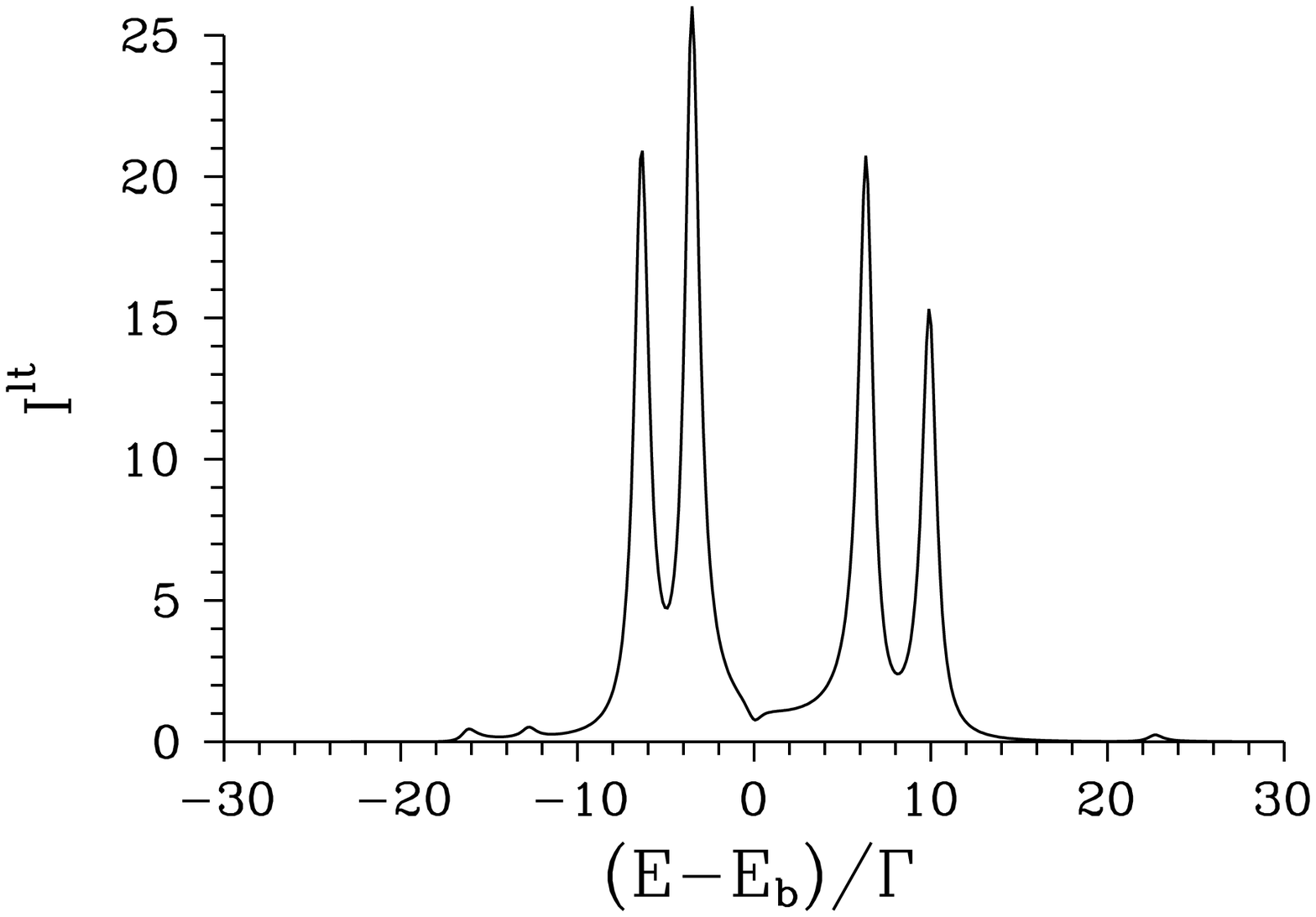}

 \caption{Long-time photoelectron ionization spectrum $ I^{\rm lt} $;
  values of parameters given in the caption to Fig.~\ref{fig3} are used.}
\label{fig9}
\end{figure}

We illustrate typical properties of the spectral entanglement
considering the system characterized by parameters mentioned in
the caption to Fig.~\ref{fig3}. In Fig.~\ref{fig8}, the spectral
density $ n_q $ of quadratic negativity is plotted in the range of
relative frequencies that covers two complex peaks occurring in
the ionization spectrum (shown in Fig.~\ref{9}). Strong spectral
correlations inside the complex peaks as well as between different
peaks are clearly visible. They mainly occur in spectral regions
where the fast intensity variations occur (compare
Figs.~\ref{fig8} and \ref{fig9}).

\begin{figure}  
 \includegraphics[scale=0.3]{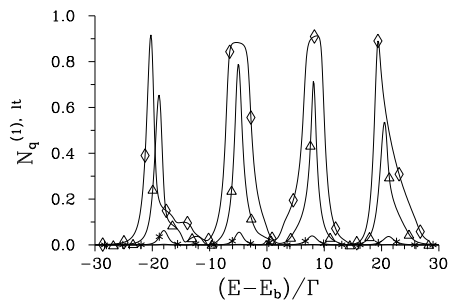}
 \caption{Long-time experimental quadratic negativity $ N_q^{(1),\rm lt} $ as a function of
  relative frequency $ (E-E_b)/\Gamma $ for $ \Delta E /\Gamma
  = 0.001 $ (solid curve), $ \Delta E /\Gamma
  = 0.005 $ (solid curve with $ \ast $), $ \Delta E /\Gamma
  = 0.025 $ (solid curve with $ \triangle $), and $ \Delta E /\Gamma
  = 0.05 $ (solid curve with $ \diamond $); $ \Gamma =
  \pi |V|^2 + \pi |J|^2 $, values of parameters given in the
  caption to Fig.~\ref{fig3} are used.}
\label{fig10}
\end{figure}

The experimental quadratic negativity $ N_q^{(1)} $ defined in
Eq.~(\ref{45}) represents the simplest experimentally accessible
characteristics. As its definition indicates, the negativity $
N_q^{(1)} $ depends on the experimental frequency resolution $
\Delta E $. It even holds that $ N_q^{(1)}(E) \rightarrow 0 $ for
$ \Delta E \rightarrow 0 $. This reflects the fact that at least a
'small' group of states $ |E\rangle $ inside the frequency
interval $ \Delta E $ is needed to 'imprint' the entanglement. The
wider the frequency interval $ \Delta E $ is, the larger are the
values of quadratic negativity $ N_q^{(1)} $. As an example, the
long-time 'distribution' of entanglement along the relative
frequency axis $ (E-E_b)/\Gamma $ for the case studied in
Fig.~\ref{fig3} is shown in Fig.~\ref{fig10}. According to
Fig.~\ref{fig10} there exist four spectral regions that
considerably contribute to the formation of entanglement. If the
frequency interval $ \Delta E $ is sufficiently wide, the maximum
attainable values of quadratic negativity $ N_q^{(1),\rm lt} $ can
be approached. The comparison of the graph in Fig.~\ref{fig10}
with that in Fig.~\ref{fig9} giving the long-time photoelectron
ionization spectrum $ I^{\rm lt} $ reveals that two spectral
regions in the middle are crucial for constituting the
entanglement between two electrons.

We note that the experimental quadratic negativity $ N_q^{(1)} $
is time-independent in the long-time limit provided that atom $ a
$ is resonantly pumped. We remind that this is not the case of
conditional long-time photoelectron ionization spectra $ I^{\rm
lt}_0 $ and $ I^{\rm lt}_1 $ obtained for atom $ a $ being in the
ground ($ |0\rangle_a $) and the excited ($ |1\rangle_a $) state,
respectively (for details, see \cite{PerinaJr2011a}).

\begin{figure}  
 (a) \includegraphics[scale=0.25]{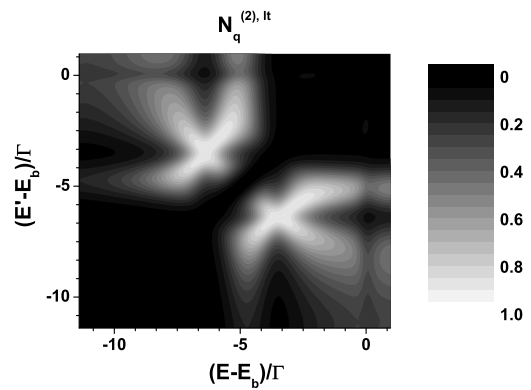}

 \vspace{7mm}
 (b) \includegraphics[scale=0.25]{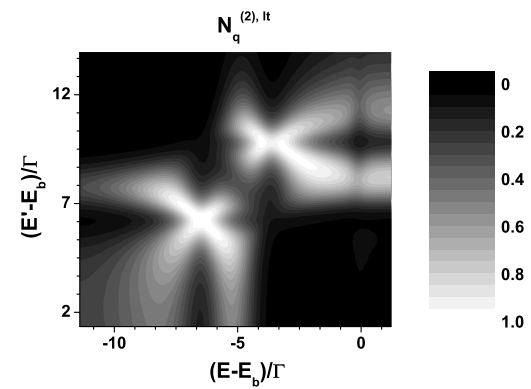}

 \caption{Topo graphs of long-time experimental quadratic negativities
  $ N_q^{(2),\rm lt} $ depending on relative frequencies $ (E-E_b)/\Gamma $
   and $ (E'-E_b)/\Gamma $ and showing the correlations (a) inside one spectral region
   and (b) between two different spectral regions; $ \Delta E /\Gamma
   = 0.001 $, values of the other parameters are written in the
  caption to Fig.~\ref{fig3}.}
\label{fig11}
\end{figure}

The spectral correlations of entanglement as theoretically
described by the density $ n_q(E,E') $ of quadratic negativity can
be experimentally revealed measuring the experimental quadratic
negativity $ N_q^{(2)}(E,E') $ introduced in Eq.~(\ref{45}). As
the considered example documents in Fig.~\ref{fig11}, two kinds of
the spectral correlations of entanglement may be distinguished.
Strong correlations are found among the frequencies $ E $ and $ E'
$ lying inside one spectral peak, but different sub-peaks [see
Fig.~\ref{fig10}(a)]. On the other hand, strong correlations occur
also for the frequencies $ E $ and $ E' $ localized inside the
neighbor spectral peaks. Here, the correlations are observed
inside the lower-frequency sub-peaks of two neighbor spectral
peaks as well as inside the upper-frequency sub-peaks of the
neighbor peaks [see Fig.~\ref{fig10}(b)]. This example illustrates
richness of the internal spectral structure of entangled stated in
the investigated system.

\section{Conclusions}

The entanglement between two electrons in an auto-ionization atom
and a neighbor two-level atom has been investigated. An expression
for the negativity of a bipartite system composed of a qubit and a
general system including both the discrete and continuum levels
has been derived. The spectral density of quadratic negativity has
been introduced to study the spectral features of entanglement. It
has allowed to decompose the overall entanglement into the
qubit-qubit entanglement of the constituting parts. Also the
concept of experimental quadratic negativities has been
introduced. It has been shown that the dipole-dipole interaction
creates the entanglement between electrons until one of them is
completely ionized. This puts restrictions to the strength of
ionization paths in the auto-ionization atom. However, the
balancing of two ionization paths in the auto-ionization atom
results in a lower ionization speed that is in favor of the
entanglement generation. Highly entangled states stable for long
times are then reached. The entanglement is spectrally
'concentrated' below the peaks of the long-time ionization
spectra. Strong correlations have been found for pairs of
frequencies localized inside one spectral peak as well as when two
frequencies have been positioned below the neighbor peaks.

\appendix

\section{Alternative derivation of the formula (\ref{32}) for negativity
$ N $}

We may conveniently decompose the functions $ d_0(E) $ and $
d_1(E) $ characterizing an ionized electron at atom $ b $ in a
suitable orthonormal basis formed by functions $ f_0(E) $ and $
f_1(E) $. In this basis, the problem of quantifying entanglement
between the two-level system $ a $ and the system $ b $ with the
continuum of states is reduced to the problem of quantifying the
entanglement in a qubit-qubit system. The appropriate basis
functions $ f_0(E) $ and $ f_1(E) $ can be constructed along the
following recipe:
\begin{eqnarray}  
 f_0(E) &=& \frac{d_0(E)}{\sqrt{b_{00}}} , \nonumber \\
 f_1(E) &=& \frac{ -b_{10}d_0(E) +
  b_{00} d_1(E) }{b_{00}b_{11} -|b_{01}|^2 };
\label{A1}
\end{eqnarray}
the coefficients $ b_{jk} $ have been defined in Eq.~(\ref{28}).
The inverse transformation to that written in Eq.~(\ref{A1}) can
be derived in the form:
\begin{eqnarray}  
 d_0(E) &=& \alpha_{00} f_0(E), \nonumber \\
 d_1(E) &=& \alpha_{10} f_0(E) + \alpha_{11} f_1(E) ;
\label{A2}
\end{eqnarray}
$ \alpha_{00} = \sqrt{b_{00}} $, $ \alpha_{10} =
b_{10}/\sqrt{b_{00}} $, and $ \alpha_{11} = (b_{00}b_{11} -
|b_{10}|^2)/b_{00} $.

Using new basis vectors $ |0\rangle\rangle_b $ and $
|1\rangle\rangle_b $ in the continuum of states at atom $ b $,
\begin{equation}   
 |j\rangle\rangle_b = \int \, dE f_j(E) |E\rangle , \hspace{5mm} j=0,1,
\label{A3}
\end{equation}
the state vector $ |\psi\rangle^{\rm lt} $ in Eq.~(\ref{5}) can be
recast into the following long-time form:
\begin{eqnarray}  
 |\psi\rangle^{lt} &=& \alpha_{00} |0\rangle_a |0\rangle\rangle_b +
  \alpha_{10} |1\rangle_a |0\rangle\rangle_b + \alpha_{11} |1\rangle_a
  |1\rangle\rangle_b .
  \nonumber \\
 & &
\label{A4}
\end{eqnarray}

The state vector $ |\psi\rangle^{\rm lt} $ can be considered as a
state of two qubits, $ a $ and $ b $. The partially transposed
statistical operator $ \varrho^{Ta} $, transposed with respect to
the indices of atom $ a $, can be written in the following matrix
form:
\begin{equation}   
 \varrho^{Ta} = \left[ \begin{array}{cccc}
   \alpha_{00}^2 & 0 & \alpha_{00}\alpha_{10} & 0 \\
   0 & 0 & \alpha_{00}\alpha_{11} & 0 \\
   \alpha_{00}\alpha_{10}^* & \alpha_{00}\alpha_{11} &
   |\alpha_{10}|^2 & \alpha_{10}\alpha_{11} \\
   0 & 0 & \alpha_{10}^*\alpha_{11} & \alpha_{11}^2 \end{array}
   \right] .
\label{A5}
\end{equation}
The secular equation for the matrix $ \varrho^{Ta} $ can be
obtained in the form $ (\lambda^2 - {\cal D})( \lambda^2 -
p\lambda + {\cal D}) = 0 $, where $ p = \alpha_{00}^2 +
\alpha_{11}^2 + |\alpha_{10}|^2 $ and $ {\cal D} $ has been
defined in Eq.~(\ref{30}). The only negative solution of the
secular equation, $ \lambda = - \sqrt{\cal D} $, gives the formula
for negativity $ N $ given in Eq.~(\ref{32}).

\acknowledgments Support by projects COST OC 09026, 1M06002 and
Operational Program Research and Development for Innovations -
European Social Fund (project CZ.1.05/2.1.00/03.0058) of the
Ministry of Education of the Czech Republic are acknowledged. Also
support by project PrF-2011-009 of Palack\'{y} University is
acknowledged.

\bibliography{luks}
\bibliographystyle{apsrev}

\end{document}